\documentclass{article}

\usepackage{graphicx}
\usepackage{mathtools}
\usepackage{amsmath}
\usepackage{booktabs}

\usepackage[top=2cm,bottom=2cm,left=2.5cm,right=2cm]{geometry}

\usepackage{lipsum}
\usepackage{graphicx} 
\usepackage{amsthm} 
\usepackage{amsmath}
\usepackage{amssymb}  
\usepackage[linesnumbered,ruled,vlined,noend]{algorithm2e}
\usepackage{enumitem}
\usepackage{tikz}
\usepackage{caption}
\usepackage{hyperref}
\usetikzlibrary{arrows.meta, positioning, shapes.geometric, calc}

\usepackage{placeins}

\newtheorem{lemma}{Lemma}
\newtheorem{theorem}{Theorem}

\newtheorem*{corollary*}{Corollary}
\newtheorem*{proposition*}{Proposition}
\newtheorem*{theorem*}{Theorem}

\title{Prism Tables: Improving Time-Memory Tradeoffs for Function Inversion\footnote{Department of Computer Science and Applied Mathematics, Weizmann Institute of Science, Rehovot, Israel. Research supported in part by a grant from the Simons Foundation Collaboration on the Theory of
Algorithmic Fairness}}
\author{Moni Naor \and Michael Sieradzki}
\date{November 30, 2026}
\begin{document}
\maketitle
\begin{abstract}
We introduce \emph{prism tables}, which improve the time-memory
tradeoff for inverting a function $f\colon[N]\to[N]$ by a factor of $\log N$.
For a random function, it gives $TM^2=O(N^2/\log N)$, which
is the
first asymptotic improvement on Hellman's classical 1980 tradeoff of $TM^2=O(N^2)$. Here $M$ counts the words of
precomputed advice about $f$ and $T$ the expected number of
online evaluations needed to find a preimage of a given challenge.
For arbitrary functions the tradeoff becomes $TM^2=O(N^3q(f)/\log N)$, where $q(f)$ is the collision probability of $f$, and $TM^3=O(N^3/\log N)$ in the worst case. 

We give rainbow tables (introduced by Oechslin in 2003) the rigorous worst-case
guarantees that Fiat and Naor established for Hellman's tables.
Rather than relying on randomness assumptions, 
the tables use explicit pairwise independent functions, stored as part of the advice, and counted as part of their storage. Following Fiat and Naor, we store images with many preimages
separately and let the chains bypass them.

Prism tables extend this rainbow construction by storing compact
information that tells the online search whether it is on the right
track or on a path leading nowhere. Concretely, we place a Bloom filter 
at $\log N$
selected columns, each including all the chain values in its column.
The filters let the search identify and discard
most wrong column (color) guesses well before the endpoint. This cuts the
expected work on incorrect guesses, including false alarms, by a
factor of $\log N$, and with it the online cost. Since the filters
have no false negatives, prism tables preserve the success probability
of the underlying rainbow tables.

The improvement
is already visible in experiments on AES-based functions with $N\approx 2^{20}$ to $2^{32}$: on average, prism tables used $1.92$–$2.95\times$ fewer online evaluations than rainbow tables at equal storage and similar success rate, with the gain growing like $\log N$.
\end{abstract}

\section{Introduction}
In the \emph{function-inversion} problem, we are given a function
$f\colon[N]\to[N]$ and a challenge $y=f(x)$, and we seek a preimage
$x'\in[N]$ such that $f(x')=y$. We assume nothing about the structure
of $f$ and access it only through oracle calls. This black-box model
captures generic attacks such as recovering a password from its hash,
where $f$ maps each candidate password to its hash, or recovering a
block-cipher key from a known plaintext, where $f(x)=E_x(P_0)$ for a
fixed plaintext $P_0$. In either setting, an attacker may be able to invest
computation before the challenge arrives to speed up the subsequent
search for a preimage. This can be modeled in two phases. During
preprocessing, an algorithm queries $f$ and stores $M$ words of
advice, each of $\log N$ bits. Once $y$ arrives, the online algorithm
uses this advice to find a preimage of $y$, making at most $T$
evaluations of $f$ in expectation.

Hellman~\cite{Hel80} initiated the study of such tradeoffs in 1980 and
gave an algorithm, known as \emph{Hellman's tables}, that builds chains
by alternating between applying $f$ and a map from images back to
inputs, with a different map for each table. It achieves
$TM^2=O(N^2)$ assuming that $f$ is a random function and that the
compositions of $f$ with these maps behave like independent random
functions.

Two central lines of work have since built on Hellman’s tables.
Fiat and Naor~\cite{FiatNaor} made Hellman's approach
rigorous and extended the algorithm to handle arbitrary functions, while Oechslin's rainbow
tables~\cite{Oec03}, a variant of Hellman's tables, became a practical choice for cryptanalysts because
they were simpler to build and faster to search.

The Fiat-Naor construction achieves $TM^2=O(N^3q(f))$ for every $f$,
where $q(f)=\Pr_{x,x'\sim{[N]}}[f(x)=f(x')]$ is the collision probability of $f$.
It recovers
Hellman's tradeoff for random functions, and gives
$TM^3=O(N^3)$ in the worst case. The rigorous construction has
since become a building block in data structures and complexity
theory. It underlies data structures that refute the strong
3SUM-Indexing conjecture~\cite{KP19,GGHPV20}, and non-uniform
circuits that beat exhaustive search for the Minimum Circuit Size
Problem and time-bounded Kolmogorov complexity~\cite{HIW24,MazorPass2024}.

Rainbow tables underlie password-recovery tools such as RainbowCrack
and Ophcrack~\cite{RainbowCrack,Ophcrack} and were used to break GSM~\cite{Meyer11}. Although salting (which hashes each password together with a random
value stored alongside its hash) reduces their effect by requiring a
separate table for each salt~\cite{MorrisThompson79},
they remain in active use today: in January 2026 Mandiant (part of Google Cloud) released
rainbow tables that recover the 56-bit DES keys of the legacy
Net-NTLMv1 protocol~\cite{Losby26}. Subsequent work made rainbow tables more
efficient through false-alarm (endpoint matches that do not lead to a preimage) detection~\cite{AJO05,ABC15}, compact
storage~\cite{AC13}, and perfect tables~\cite{AJO08}. 
These refinements
improved the constants while all retain the rainbow tables tradeoff of $TM^2=O(N^2)$
for random functions. As with Hellman's tables, Oechslin's analysis
relies on heuristic randomness assumptions.

\paragraph{Contributions.}
First, we unify these two lines of work by giving \emph{rigorous rainbow tables}, a version of Oechslin's construction that attains the Fiat-Naor tradeoffs for arbitrary functions
(Theorem~\ref{thm:rainbow}).
Second, and as our main
result, we show that the expected online time of function inversion can
be reduced by a factor of $\log N$ at the same memory and success
probability, using \emph{prism tables}, which are built on our rigorous
rainbow tables construction.

\begin{theorem}[Main result, informal; see Theorem~\ref{thm:prism}]
\label{thm:main-informal}
For every $f\colon[N]\to[N]$ and chains of length $t\ge7\log N$, prism
tables store $M$ words of advice and invert every fixed challenge in the
image of $f$ with probability at least $3/5$ over the preprocessing,
using an expected $T$ evaluations of $f$, where
\[
TM^2=O\!\left(\frac{N^3q(f)}{\log N}\right)
\]
and, except with negligible probability,
\[
TM^3=O\!\left(\frac{N^3}{\log N}\right).
\]
\end{theorem}
For a random function,
Theorem~\ref{thm:main-informal} improves the $TM^2=O(N^2)$ tradeoff of
Hellman's and rainbow tables to $TM^2=O(N^2/\log N)$, which is the first asymptotic improvement to the
tradeoff outside the regime $T=O(\log N/\log\log N)$, where Kuszmaul and Kuszmaul~\cite{KK26} recently achieved $TM=O(N)$. For arbitrary functions, it improves the Fiat-Naor
worst-case tradeoff from $TM^3=O(N^3)$ to $TM^3=O(N^3/\log N)$.

Experiments on AES-derived functions with $N\approx2^{20},\ldots,2^{32}$
confirm that this gain already appears at practical sizes: at equal
storage and similar success rate, prism tables used $1.92$--$2.95\times$ fewer
evaluations of $f$ and $1.83$--$2.80\times$ less CPU time than rainbow
tables, with a gain that grows like $\log N$, as the analysis predicts
(Section~\ref{sec:experiments}).

\vspace{0.1in}
\noindent
{\bf Note:} Concurrently and independently of this work, Dinur, Golovnev, Yan and
Zhu~\cite{DGYZ26} give a rigorous rainbow-table construction, similar to the one in Section~\ref{sec:rigorous-raibow-tables}, emphasizing function reporting (we only describe obtaining a single preimage). This is then used for 3SUM-Indexing Reporting, Range Reporting, Approximate Nearest Neighbor, and other data structure problems.  
They do not consider prism tables (Section~\ref{sec:prism}). 

\subsection{From Rainbow Tables to Prism Tables}
\begin{figure}
    \centering
    \includegraphics[width=0.85\linewidth]{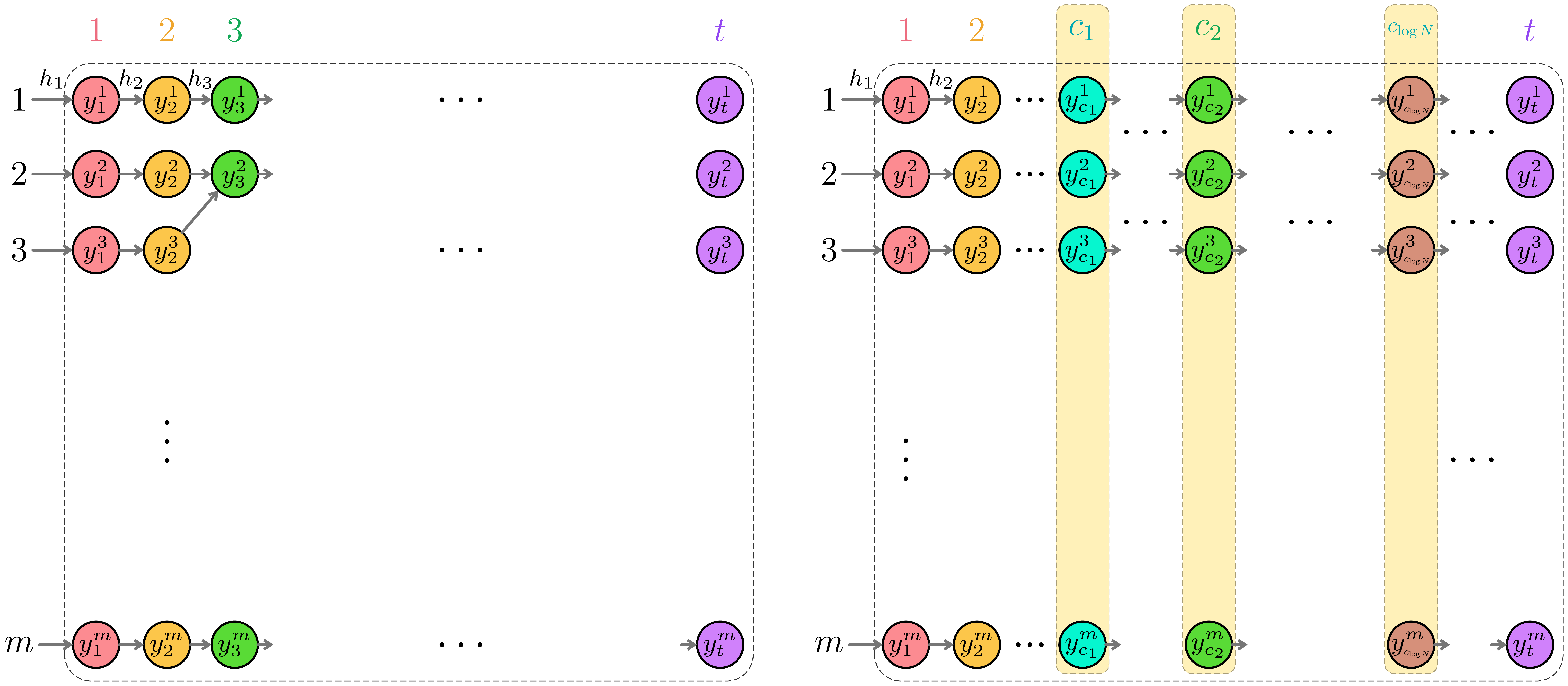}
      \caption{Left: a rigorous rainbow table. Right: a prism table, with Bloom filters at the
checkpoint columns $c_1,\ldots,c_{\log N}$.}
    \label{fig:rainbow-vs-prism}
\end{figure}

A rainbow table consists of $m$ chains of length $t$ whose links are
images of $f$. To move from one column to the next, a chain uses the
new column's \emph{transition function}, or \emph{color}, $h_s=f\circ g_s$, which maps
the last image to an input of $f$ and then applies $f$ to obtain
the next image (Fig.~\ref{fig:rainbow-vs-prism}, left). The table then stores only the start and endpoint of each chain. Oechslin's analysis treats $f$
and the colors as random functions. In our rigorous version, each $g_s$
is drawn from a pairwise independent family, and its description is
stored in the advice and counted in $M$, so the analysis holds for every
fixed $f$, with all randomness coming from the sampled $g_s$. For the worst-case bound we follow Fiat and Naor and bypass the
\emph{heavy} images, those with many preimages, at which chains are
especially likely to \emph{merge}, that is, when they reach the same image in
the same column and thus share the rest of their path from then on.

During the online phase, if the algorithm
knew the column containing $y$, it could apply the sequence of transition functions to follow the chain containing $y$ to its endpoint. Then it could look up that endpoint and reconstruct the matching chain from its starting point to reach the link
preceding $y$, which gives a preimage of $y$.
However, the online algorithm does not know which column contains $y$.
It therefore \emph{guesses} a column (the color) that contains $y$ and makes a \emph{candidate walk} for the remaining steps to the endpoint column, where it can first check the walk
against the table. An incorrect guess usually misses the stored endpoints, as it applies the incorrect sequence of transition functions, but if its walk
merges with a stored chain, the match is a \emph{false alarm} that
triggers a needless reconstruction. Trying all columns costs about
$t^2/2$ evaluations of $f$ just to reach the endpoint column, while a
correct guess takes only $O(t)$ steps. Most of the online work therefore goes into
rejecting incorrect guesses~\cite{BBS06}. Can an incorrect guess be recognized sooner?

Prism tables can detect most incorrect column (color) guesses long before
their walks reach the endpoint column by adding \emph{membership
checkpoints}%
\footnote{Not to be confused with the checkpoints of Avoine, Junod and
Oechslin~\cite{AJO05}, which are consulted only after an endpoint
match. In contrast, our checkpoints are consulted during the walk and can reject a
guess along it.} to a rainbow table: at $\log N$ evenly spaced columns, a
Bloom filter\footnote{Bloom filters~\cite{Bloom70} are a data structure used for set membership tests. They are fast and have very good space efficiency, at the price of adding a certain false positive probability, see Mitzenmacher-Upfal~\cite{MitzenmacherU17} 
Section 5.5.3.} with false-positive probability at most
$1/2$ stores all chain points in that column (Fig.~\ref{fig:rainbow-vs-prism}, right). During the online phase,
whenever a candidate walk reaches a checkpoint
column, we query that column's filter with the walk's current point. A
negative answer means that no stored chain passes through this point in
this column, so we discard the guess, while a positive answer lets the
walk continue to the next checkpoint, where the test is repeated, or to
the endpoint column, where the search proceeds as in a standard rainbow
table. Since the Bloom filters have no false
negatives, they never discard a correct guess, so the success
probability of the underlying table is unchanged.

The savings come from shortening the walks of incorrect guesses. Until it
merges with a stored chain, an incorrect candidate walk passes each filter
with probability at most $1/2$, so it travels about two checkpoint
intervals on average. Its expected length is therefore about
$2t/\log N$ steps so rejecting all incorrect guesses costs $O(t^2/\log N)$ evaluations per table instead of about $t^2/2$.
The shorter walk also gives fewer chances to merge with a stored
chain and cause a false alarm. Each filter stores the $m$ points of one column and needs about $2m$
bits, while the starting points and endpoints of the $m$ chains in the
same table occupy $2m\log N$ bits. We can therefore afford $\log N$
membership checkpoints per table while increasing its space by only
a constant factor. 

Prism's checkpoints are thus a layer on top of the chains rather than a new way
of building them. And although we prove our bounds for rainbow tables,
we expect the same layer to give similar savings for many chain-based
constructions, such as perfect tables~\cite{AJO08} and table compression~\cite{AC13}. If we use such compression, each chain takes
fewer bits while the filters do not shrink, so they take a larger share
of the memory, which lowers the constant of the improvement.

\subsection{Related Work}
The best lower bound for arbitrary algorithms is still Yao's
$TM=\Omega(N)$ from 1990~\cite{Yao90}, which leaves a wide
gap to the $TM^2=O(N^2)$ of Hellman's tables.
Corrigan-Gibbs and Kogan~\cite{CK19} explained this gap by showing that
sufficiently improving Yao's bound, even for strongly non-adaptive algorithms, whose queries depend only on $y$, would imply new circuit lower bounds. Lower bounds are
therefore known only for restricted classes of algorithms. Stronger bounds
hold for inverters with linear advice and for non-adaptive inverters with
affine decoders~\cite{CHM20}, and strongly non-adaptive inverters need $TM=\Omega(N\log T)$~\cite{GGK24}.
Non-adaptive guess-and-check inverters, which output one of their queries,
have matching upper and lower bounds~\cite{GGPS23}, and the same work
improved the Fiat-Naor worst case to $T^2M^2=\widetilde{O}(N^3)$ when
$T>M$ by rebuilding the heavy-image table online.
Barkan, Biham and Shamir~\cite{BBS06} gave
a unified model capturing Hellman's tables and rainbow tables, as well as
other potential variants, and showed that they are optimal up to
logarithmic factors within the model. Their model, however, assumes that
the online algorithm follows each incorrect column guess all the way to the
endpoint column, so prism tables, which reject most incorrect guesses at
intermediate checkpoints, fall outside it.

More broadly, a long
line of work builds on the Fiat-Naor scheme by applying it to data
structure problems, including
3SUM-Indexing~\cite{KP19,GGHPV20,DG25}, preprocessed 3SUM with
subset queries~\cite{KirkpatrickKMW26,CKK26}, searching
in implicit sets~\cite{ACDI26} and text indexing with
mismatches~\cite{BBM26}. Other work extends the scheme itself, to point
updates and the small-time regime~\cite{KK26} and to reporting all
preimages~\cite{Van26,DGYZ26}.

On the practical side, Hong and Moon~\cite{HM13} compare Hellman's
tables, distinguished-point tables and rainbow tables. The closest
predecessors to prism tables are the checkpoints of Avoine, Junod and
Oechslin~\cite{AJO05} and the fingerprints of Avoine, Bourgeois and
Carpent~\cite{ABC15}, which store a few bits per chain at selected
columns to avoid false alarms, but since the online algorithm cannot tell
which chain it is on, it can compare these bits only at the end of its
walk. Prism tables instead check each checkpoint at its own column, which
reduces both the false alarms and the online walks.

\subsection{Organization}
Section~\ref{sec:prelim} describes the model, the parameters and our
notation, explains how we build the column functions, and recalls two
standard tools that our constructions use, Bloom filters and the
Misra-Gries algorithm.
Section~\ref{sec:rigorous-raibow-tables} gives our rigorous rainbow
tables, first the basic construction with $TM^2=O(N^3q(f))$
(Section~\ref{sec:light}) and then the full construction, which
bypasses heavy images to obtain $TM^3=O(N^3)$
(Section~\ref{sec:heavy}). 
Section~\ref{sec:prism} builds prism tables by
adding membership checkpoints to these constructions and shows
that they reduce the expected online time by a factor of $\log N$
while increasing memory by only a constant factor, proving
Theorem~\ref{thm:prism}. Finally, Section~\ref{sec:experiments}
reports experiments comparing rainbow and prism tables at equal storage and similar
success probability, showing that the improvement is already
visible at practical sizes.

\section{Model and Preliminaries}\label{sec:prelim}

A preprocessing algorithm with oracle access to $f\colon[N]\to[N]$
computes an advice string of $M$ words with a word size of $\log N$ bits. Given the preprocessed advice and a challenge
$y=f(x)$, an online algorithm makes oracle calls to $f$ and must output
some $x'$ such that $f(x')=y$ with constant probability over the
randomness of the preprocessing algorithm. We
let $T$ be the largest expected number of evaluations of $f$ it makes
on any challenge.

Let $I(y)\coloneqq|\{x\in[N]:f(x)=y\}|$ be the number of preimages
of $y$. The collision probability of $f$ is
\[
q=q(f)\coloneqq
\Pr[f(x)=f(x')]
=\frac{1}{N^2}\sum_{y=1}^{N}I(y)^2,
\]
where $x$ and $x'$ are sampled independently and uniformly from $[N]$.

For a set of images $A\subseteq f([N])$, let $q^\ast=q^\ast_A(f)$
be the probability that two such inputs collide at an image
outside $A$:
\[
q^\ast
=\Pr[f(x)=f(x'),\,f(x)\notin A]
=\frac{1}{N^2}\sum_{y\in[N]\setminus A}I(y)^2.
\]
In the full construction of Section~\ref{sec:heavy}, we choose $A$ to contain the images
with many preimages, which we call \emph{heavy}. Unlike $q$, which is
at least $1/N$, $q^\ast$ can be smaller than $1/N$,
and we state our bounds in terms of $q^\ast$ for $q^\ast\ge1/N$,
while for smaller $q^\ast$ they hold with $1/N$ in place of
$q^\ast$ (Section~\ref{sec:heavy}).

\paragraph{Column functions.}\label{sec:column-functions}
We assume that $N$ is a prime power\footnote{Otherwise, we take a prime
$p$ with $N\le p<2N$, and invert
the extension of $f$ to $[p]$ that maps each $z>N$ to itself without an
oracle call. The challenges and their preimages remain in $[N]$, and
the collision probability of the extension is at most
$q(f)+1/p\le2q(f)$, so our bounds change only by constant factors.}
and identify $[N]$ with the field $\mathbb{F}_N$. A column function is
$g(y)=\alpha y+\beta$, with $\alpha$ and $\beta$ chosen independently
and uniformly from $\mathbb{F}_N$. In the full construction of
Section~\ref{sec:heavy}, its input is a pair $(y,a)$, with $a\le t<N$
also viewed as an element of $\mathbb{F}_N$, and
$g(y,a)=\alpha y+\gamma a+\beta$ with a third uniform coefficient
$\gamma$. Such a $g$ is \emph{pairwise independent}, and it is
described by $2\log N$ or $3\log N$ bits, respectively.

Our Prism tables use standard Bloom filters~\cite{Bloom70}: an approximate set membership data structure for
storing $m$ elements (from the universe of images).
The filter has no false negatives, and
has false-positive probability at most $1/2$. More specifically, for any set of elements $S$ and for any element $w \not\in S$ the probability over the choice of the hash functions of the Bloom filter that the Bloom filter returns yes is bounded by $1/2$. Note that there are stronger definitions for Bloom filters in adversarial environments~\cite{NaorY19,NaorO22,LotanN25}, but the analysis in Section~\ref{sec:analysis} will be based on the standard non-adversarial definition. 

Under the standard analysis of Bloom filters the storage size is $1/\ln 2 \cdot m \approx 1.44 m$. But this assumes complete independence of the hash functions. Under pairwise independence we get that we need no more than 2 bits per element, plus $O(1)$ words for the hash function.
There are even better constructions such as Ribbon filters~\cite{Ribbon2022} which approach 1 bit per element and cuckoo filters~\cite{FANAKM14}, which beat Bloom filters at smaller false-positive rates. 

The rainbow tables construction of Section~\ref{sec:heavy}, and hence the prism
tables built on it, uses in its preprocessing phase the Misra-Gries
algorithm, a streaming algorithm for finding frequent items (aka ``heavy hitters") using a small, fixed amount of working memory.
Given a stream of length $ck$, it keeps every item that appeared at least $c$ times while storing at most $k$ (item, counter) pairs. See~\cite{CormodeH09} for a survey. 

\paragraph{Notations.}
We index tables by $i\in[\ell]$, chains by $j\in[m]$ and columns by
$s\in\{0,\ldots,t\}$, writing $y_s^{j,i}$ for the point of chain $j$ of
table $i$ in column $s$, and $h_s^{(i)}$ for the transition of column
$s$ in table $i$. We drop the table index when we discuss a single
table, writing $y_s^j$ and $h_s$, and the chain index for a walk whose
chain is unknown, such as a candidate walk from $y$, whose point in
column $s$ we write $y_s$.

We write $\log$ for $\log_2$ rounded up to an integer, so that an
element of $[N]$ takes $\log N$ bits, and we assume throughout that
$7\le t\le N/6$ and that $t^2q<1/2$ (with $q^\ast$ in place of $q$ in
Section~\ref{sec:heavy}). The condition $t^2q<1/2$ loses nothing, since otherwise the time
bound $\ell t^2$ of Section~\ref{sec:light}, with $\ell>2Nq$ tables,
would exceed the $N$ evaluations of exhaustive search. 
\begin{center}
\begin{tabular}{cl}
\toprule
\(N\)    & size of the domain, \(f\colon[N]\to[N]\) \\
\(\ell\) & number of rainbow tables \\
\(m\)    & chains per table \\
\(t\)    & chain length \\
\(M\)    & total memory words of $\log N$ bit the preprocessing algorithm stores \\
\(T\)    & expected number of $f$ evaluations by the online algorithm \\
\bottomrule
\end{tabular}
\end{center}

\section{Rigorous Rainbow Tables}\label{sec:rigorous-raibow-tables}

Section~\ref{sec:light} gives the basic construction and proves its
success probability and its tradeoff $TM^2=O(N^3q(f))$, and
Section~\ref{sec:heavy} extends it to the full construction, which
bypasses heavy images to obtain the worst-case tradeoff
$TM^3=O(N^3)$.

\begin{theorem}[Rigorous rainbow tables]\label{thm:rainbow}
Let $f\colon[N]\to[N]$.
\begin{enumerate}
\item[(i)] The rainbow tables of Section~\ref{sec:light} invert each
fixed challenge $y=f(x)$ with probability at least $3/5$ over the
preprocessing and achieve the tradeoff
\[
TM^2=O\bigl(N^3q(f)\bigr).
\]
\item[(ii)] The rainbow tables of Section~\ref{sec:heavy}, which bypass
the images stored in $A$, invert each fixed challenge $y=f(x)$ with
probability at least $3/5$ over the preprocessing and achieve the
tradeoff
\[
TM^2=O\bigl(N^3q^\ast_A(f)\bigr).
\]
With probability at least $1-N^{-2}$ over the construction of $A$, we
have $q^\ast_A(f)<8/M$, so the worst-case tradeoff becomes
\[
TM^3=O(N^3).
\]
\end{enumerate}
\end{theorem}

\subsection{The Basic Construction} \label{sec:light}
\paragraph{Preprocessing.}

We first describe the construction of a single rainbow table; the preprocessing
algorithm then builds \(\ell\) such tables, using independent randomness for each.

Since the preprocessing chooses $m$ and $\ell$ from $q$, it must first
estimate $q$, which it can do cheaply only up to a constant factor. As
in~\cite{FiatNaor}, we can obtain a slightly pessimistic tradeoff by
taking a slightly higher value for $q$, since the analysis below uses $q$
only as an upper bound on the probability that two chains meet in a column.

\textbf{A single rainbow table.}
Preprocessing samples the column functions $g_1,\ldots,g_t$
independently at random as in Section~\ref{sec:column-functions} and
sets:
\[
h_s\coloneqq f\circ g_s.
\]
A rainbow table consists of $m=\lceil1/(tq)\rceil$ \emph{chains} of
length $t$, where chain $j\in[m]$ starts at the \emph{starting point}
$y_0^j=j$ and follows $y_s^j=h_s(y_{s-1}^j)$ for each $s\in[t]$.
The resulting chains have the form
\[
C_j:\quad
y_0^j \xrightarrow{\;h_1\;}
y_1^j \xrightarrow{\;h_2\;}
\cdots \xrightarrow{\;h_t\;}
y_t^j,
\qquad j\in[m],
\]
where $y_s^j$ lies in column $s$ and $y_t^j$ is the
\emph{endpoint} of chain $j$ (Fig.~\ref{fig:rainbow-vs-prism}, left).

Two chains $C_j$ and $C_{j'}$ \emph{merge} if they reach the same image
in some column. We say they \emph{merge in column $s$} if $s$ is the
smallest column index for which $y_s^j=y_s^{j'}$. From that column onward, each step applies the same
function to the same image in both chains, so they follow the same
sequence of images all the way to the endpoint. The same image
appearing in different columns does not constitute a merge, since
the two chains continue with different column functions and may
follow different paths.

For two chains that have not yet merged, their points $u,v$ in
column $s-1$ are distinct, and thus pairwise independence of the function $g_s$ makes $g_s(u)$, $g_s(v)$ independent
and uniform on $[N]$.
The two
chains therefore merge in column $s$ with probability
\[
\Pr[h_s(u)=h_s(v)]=q(f).
\]

\noindent
\textbf{Using $\ell$ rainbow tables.}
To increase the probability that a fixed challenge appears in a table,
we repeat the construction $\ell\coloneqq\lceil2Nq+1\rceil$ times,
independently sampling fresh column functions for each table, and
denote the resulting tables by $\mathcal{T}_1,\ldots,\mathcal{T}_\ell$.

A chain $C_j$ \emph{covers} $y$, written $y\in C_j$, if $y_s^j=y$ for
some column $s\in[t]$, and a table \emph{covers} $y$, written
$y\in\mathcal{T}_i$, if one of its chains does.

For each table, the
preprocessing stores only the starting point and endpoint of each
chain as the pair $(y_0^j,y_t^j)$, together with descriptions of that
table's sampled functions $g_1,\ldots,g_t$.

\paragraph{Online phase.}
As explained in the introduction, the online algorithm does not know
the column in which its challenge $y$ appears, so it guesses it. For a
guess $r$, it develops a \emph{candidate chain} from $y$ by applying
$h_{r+1},\ldots,h_t$ and compares the resulting endpoint with the
stored endpoints of each table, ruling out the guess for a table if
none matches. Otherwise, it rebuilds each stored chain with this
endpoint from its starting point through column $r-1$, and if the
image $y_{r-1}$ reached there satisfies $h_r(y_{r-1})=y$, it returns
$x=g_r(y_{r-1})$, which is a preimage since $f(x)=h_r(y_{r-1})=y$.

The algorithm tries $r=t,t-1,\ldots,1$ in this order, testing each
guess in all $\ell$ tables before moving to the next. It stops as soon as it
finds a preimage and returns $\bot$ if every guess fails
(Algorithm~\ref{alg:online-rainbow}).

\begin{algorithm}
\caption{Online inversion with rainbow tables}
\label{alg:online-rainbow}
\KwIn{$f$, $y\in[N]$, and the tables $\mathcal{T}_1,\ldots,\mathcal{T}_\ell$}
\For{$r=t,t-1,\ldots,1$}{
    \For{$i=1,\ldots,\ell$}{
        $z\gets (h_t^{(i)}\circ\cdots\circ h_{r+1}^{(i)})(y)$\;
        \ForEach{chain $C_j$ in $\mathcal{T}_i$ with endpoint $z$}{
            $u\gets (h_{r-1}^{(i)}\circ\cdots\circ h_1^{(i)})(y_0^j)$\;
            \lIf{$h_r^{(i)}(u)=y$}{\Return{$g_r^{(i)}(u)$}}
        }
    }
}
\Return{$\bot$}\;
\end{algorithm}

\subsubsection{Analysis}

Fix $x\in[N]$ and let the challenge $y=f(x)$, with all probabilities
below taken over the independent choices of the column functions
$g_s$ in each table. We first give a lower bound on the probability
that a single table covers $y$ and then use independence between
the tables to show that at least one of them covers $y$ with
probability at least $3/5$.

\begin{lemma}[Single table coverage]\label{thm:pr y in table}
For every $i\in[\ell]$,
\[
\Pr[y\in\mathcal{T}_i]
\ge \frac{1-\frac{t}{2N}}{2Nq+1}.
\]
\end{lemma}

\begin{proof}
Fix a table $\mathcal{T}_i$ with $m=\lceil1/(tq)\rceil$ chains
$C_1,\ldots,C_m$. Let
$X=\sum_{j=1}^m\mathbf{1}_{\{y\in C_j\}}$ count the chains covering
$y$, and write $p_y=\Pr[y\in C_j]$ for the probability that a fixed
chain covers $y$. Since the table covers $y$ exactly when $X>0$,
the second-moment method gives
\[
\Pr[y\in\mathcal{T}_i]=\Pr[X>0]
\ge \frac{\mathbb{E}[X]^2}{\mathbb{E}[X^2]}.
\]
Each chain has the same coverage probability $p_y$, so
$\mathbb{E}[X]=mp_y$, and expanding $X^2$ gives
\[
\mathbb{E}[X^2]
=mp_y+\sum_{j\ne j'}\Pr[y\in C_j,\,y\in C_{j'}].
\]
We next derive a lower bound on $p_y$ and an upper bound on the
probability that two distinct chains both cover $y$. The second-moment
argument involves at most two chains at a time, which is why we can
use a pairwise independent family for the independently sampled
column functions.

For a single chain, each column uses a fresh function $g_s$, so the
input to $f$ is uniform on $[N]$, independently of the preceding
columns. Each step therefore produces $y$ with probability $I(y)/N$,
giving
\[
p_y=1-\left(1-\frac{I(y)}{N}\right)^t
\ge 1-e^{-t/N}
\ge (1-\tfrac{t}{2N})\frac{t}{N},
\]
where we use $I(y)\ge1$ and the inequality
$1-e^{-z}\ge z-z^2/2$ for $z\ge0$.

To give an upper bound on $\Pr[y\in C_j,\,y\in C_{j'}]$ for
$j\ne j'$, we consider two cases. If $y$ appears after the chains
have merged, a single occurrence on their shared path is enough
for both chains to cover $y$. Otherwise, each chain must encounter
$y$ along its own path before or at their first merge, with this
case also including chains that never merge.

For the first case, once the chains merge, the remaining columns use
fresh independent functions, so the probability that their shared
path subsequently contains $y$ is at most $p_y$. Before the merge,
each column gives probability $q$ of merging, so a union bound over
the $t$ columns gives
$\Pr[C_j\text{ and }C_{j'}\text{ merge}]\le tq$, and
the contribution of this case is therefore at most $p_y tq$.

For the second case, pairwise independence of each fresh $g_s$
makes the chains behave as two independent chains up to and
including their first meeting. The probability that both
encounter $y$ by then is at most the probability that two
independent chains cover $y$ over all $t$ steps, namely $p_y^2$.
Combining the two cases gives
\[
\Pr[y\in C_j,\,y\in C_{j'}]
\le p_y^2+p_y tq
= p_y(p_y+tq).
\]

Returning to the second-moment calculation, we obtain
$\mathbb{E}[X^2]\le mp_y+m(m-1)p_y(p_y+tq)
\le mp_y(1+mp_y+mtq)$.
Together with $\mathbb{E}[X]^2=m^2p_y^2$, this gives
\[
\Pr[y\in\mathcal{T}_i]
\ge \frac{mp_y}{1+mtq+mp_y}
\ge \frac{p_y}{2tq+p_y}
\ge \frac{1-t/(2N)}{2Nq+1},
\]
where the second inequality uses $mtq\ge1$ so $1+mtq\le2mtq$, while the last follows from
the single-chain lower bound on $p_y$, since $p_y/(2tq+p_y)$ is
increasing in $p_y$.
\end{proof}

\begin{lemma}[Coverage]
Let $\ell=\lceil 2Nq+1\rceil$. Then the $\ell$ independently
constructed tables satisfy
\[
\Pr\left[y\in\bigcup_{i=1}^{\ell}\mathcal{T}_i\right]
\ge \frac35.
\]
\end{lemma}

\begin{proof}
By the single-table coverage bound, each
table covers $y$ with probability at least
$(1-t/(2N))/(2Nq+1)$. Since the tables use independent
randomness, the probability that all of them miss $y$ is
at most
\[
\Pr\left[y\notin\bigcup_{i=1}^{\ell}\mathcal{T}_i\right]
\le \left(1-\frac{1-t/(2N)}{2Nq+1}\right)^\ell
\le e^{-1+t/(2N)}
\le e^{-11/12}
<\frac25,
\]
where we use $\ell=\lceil 2Nq+1\rceil$ and $t\le N/6$.
Thus at least one table covers $y$ with probability greater
than $3/5$. 
\end{proof}

\paragraph{Time:}
Fix the challenge $y$. When the online algorithm guesses that $y$
appears in column $r$, it extends a candidate chain through
the remaining $t-r$ columns in each table. Trying all guesses
$r=1,\ldots,t$ in each of the $\ell$ tables, therefore costs at most
$\ell t^2/2$ transitions.

We also need to account for false alarms. Recall that each false alarm
corresponds to a stored chain that merges with the candidate chain
after the guessed column.

At each column, the probability that the candidate chain merges with
any given stored chain is at most $q$. For a guess that $y$ appears
in column $r$, a union bound over the remaining $t-r$ columns gives
a false-alarm probability of at most $(t-r)q$ with each stored chain.
Each false alarm costs at most another $r$ evaluations of $f$ to check, so
summing over all guesses, all $m$ chains in each table, and all $\ell$
tables gives a total expected cost of at most
\[
\ell mq\sum_{r=1}^{t}r(t-r)
\le \frac{\ell mq\,t^3}{6}
\le \frac{\ell t^2}{3},
\]
since $t^2q<1/2$ (Section~\ref{sec:prelim}) and $m=\lceil 1/(tq)\rceil$ give $mtq\le1+1/t$.
Finally, once a correct match is found, reconstructing its
chain and recovering a preimage costs at most $t$ additional
evaluations. Combining these costs gives
\[
\mathbb{E}[T]
\le \frac{\ell t^2}{2}
+\frac{\ell t^2}{3}+t
\le \ell t^2.
\]

Although $T$ counts only evaluations of $f$, the remaining online work
does not change its order: each evaluation of $f$ comes with one
evaluation of a column function, which takes $O(1)$ field operations,
and each guess makes one endpoint lookup per table, so counting these
as well, the time remains $O(\ell t^2)$.

The preprocessing, which the tradeoff does not count,
evaluates $f$ once at each of the $\ell mt$ points of the chains, which
is $O(N)$.

\paragraph{Memory.}
Each table uses $2m$ words to store the starting points and
endpoints of its $m$ chains and $2t$ words to describe its
$t$ column functions. Since $t^2q<1/2$, we have $m\ge 1/(tq)>2t$, so the total memory for
$\ell$ tables is
\[
M=2\ell(m+t)\le 4\ell m.
\]

\paragraph{Time-memory tradeoff.}
Using $T=O(\ell t^2)$ for the expected online time and
$M=O(m\ell)$ words of advice, we obtain
\[
TM^2=O(m^2\ell^3t^2)= O\left((mt)^2N^3q^3\right)
     = O(N^3q),
\]
using $\ell=O(Nq)$ for the first inequality and $mt=O(1/q)$ for the second.

\subsection{The Full Rainbow Tables Scheme} \label{sec:heavy}

In a rainbow table, merges reduce the number of distinct images covered by the table and hence the success probability of the scheme. 
The probability of a merge of two fixed chains at a given column equals the probability that two independent uniform domain inputs collide under $f$, so images with many preimages can dramatically increase the merging probability. 
With enough memory words, however, we can afford to store some of these images to reduce the effective merging probability by bypassing them. 

We keep these images in a table $A$, each with one of its preimages,
so that each entry takes $2$ words and giving $A$ half of the memory
allows it to hold up to $M/4$ images. We write $y\in A$ when $A$
stores $y$, treating $A$ as a set of images as in the definition of
$q^\ast_A$. We call an image \emph{heavy} if it has at least $8N/M$
preimages and show below that, with high probability, $A$ contains
every heavy image. Since the size of $A$ depends only on $M$, we can build $A$ before the
rainbow tables and then choose their parameters from $q^\ast_A$, which
can be estimated as in Section~\ref{sec:light} by counting only
collisions at images outside $A$.

As we develop each chain, we check whether the image produced in
the current column lies in $A$. If it does, we discard it and keep
the last (non-heavy) image $y$ the chain accepted, so that the chain can try
again in the next column using a fresh transition function. Keeping $y$ alone could
cause a merge if another chain accepts $y$ in the same column.
We therefore record each chain point as a pair
$(y,a)\in[N]\times\{0,\ldots,t\}$, where $a$ counts consecutive
rejected steps and returns to zero whenever the chain accepts a new
image.

By incorporating the bypassing of heavy images into a modified
transition $h^\ast$, we can build and search the rainbow tables
just as in Section~\ref{sec:light}, using $h^\ast$ in place of $h$.
We show that under $h^\ast$ two chains can merge only at a
non-heavy image, and that the choice of parameters and the analysis
carry over with the resulting merging probability $q^\ast$ in
place of $q$.

\textbf{A simple example.} We show how a single image can cause
chains to merge rapidly in the basic construction and how bypassing
that image brings the merging probability back to the same order
as for a random function. Suppose $f$ maps half of its inputs to
a fixed image $w$ and maps each remaining input independently and
uniformly to $[N]$. Each chain then reaches $w$ with probability
about $1/2$ in a fresh column, so $q(f)\ge 1/4$ and about half of
the chains merge at each column. Taking $A=\{w\}$, a chain
at $(y,0)$ whose next column proposes $w$ rejects it and moves to
$(y,1)$, and it keeps raising the counter until a column proposes
some $z\ne w$, which it accepts as $(z,0)$. Now two distinct chain
points can then merge only through the random half of $f$, so
$q^\ast_A(f)$ is of order $1/N$, as for a random function.

\paragraph{Preprocessing.}
We begin by constructing the table $A$. Set $c=12\ln N$ and
sample $cM/4$ independent uniform inputs from $[N]$, processing their images using Misra-Gries (see Section~\ref{sec:prelim}) with $M/4$ counters, each
capped at $c$ and kept together with the sampled input that produced
its image. We retain the images whose counters remain at the end of
this process, together with their stored inputs, so that $A$ holds at
most $M/4$ images, each with a preimage.

Chain $j$ now starts at $(j,0)$ and uses links
of the form $(y,a)$, where $y$ is the last accepted image and $a$ counts
retries when an image falls in $A$. Each column function $g_s$ now
takes $(y,a)$ as input and we sample these functions independently at random as defined in Section~\ref{sec:column-functions}. We then use
$h_s=f\circ g_s$ to define the transition function $h_s^\ast$ below.

\paragraph{Defining $h^\ast$.}
Fix a chain and let $(y,a)$ be its point in column $s-1$. To transition to column $s$ we evaluate $z=h_s(y,a)$ and set:
\[
h_s^\ast(y,a)=
\begin{cases}
(z,0),   & z\notin A,\\
(y,a+1), & z\in A.
\end{cases}
\]

Under $h_s^\ast$, an image in $A$ cannot cause two chains to merge.
To see why, suppose the two chains enter column $s$ at distinct points
$(u,a)$ and $(v,b)$. If both attempts land in $A$, the chains move
to $(u,a+1)$ and $(v,b+1)$, which remain distinct. If only one
attempt lands in $A$, that chain leaves the column with a positive
counter, while the other resets its counter to zero. The chains
therefore merge exactly when both attempts produce the same image
outside $A$.

The table now stores the start-endpoint pairs, with each endpoint stored as
the full pair $(y_t^j,a_t^j)$. At each column, a point $(y,0)$ records a newly accepted image $y$, while
a point $(y,a)$ with $a>0$ records a rejected attempt and does not constitute a new image occurrence.
\paragraph{Parameters.}
For a fixed table $A$, the analogue of $q$ from Section~\ref{sec:light}
is $q^\ast=q^\ast_A(f)$ defined in Section~\ref{sec:prelim}: for distinct
chain points $(u,a)$ and $(v,b)$, pairwise independence of $g_s$ gives
\[
\Pr\bigl[h_s^\ast(u,a)=h_s^\ast(v,b)\bigr]
=\sum_{z\notin A}\left(\frac{I(z)}{N}\right)^2
=q^\ast.
\]
We use the parameters of
Section~\ref{sec:light} with $q^\ast$ in place of $q$, taking
$m=\lceil1/(tq^\ast)\rceil$ chains per table and
$\ell=\lceil 2Nq^\ast+1\rceil$ independent tables. For the rest of this section we assume $q^\ast\ge 1/N$.\footnote{If $q^\ast<1/N$, we build the tables with $1/N$ in place of
$q^\ast$. Since the analysis below uses $q^\ast$ only as an upper bound
on the merging probability, it carries over and gives
$TM^2=O(N^2)$.}

\paragraph{Online phase.}
Given a challenge $y$, we first check whether $y\in A$ and, if so,
return its stored preimage. Otherwise, we search the rainbow tables
as in Section~\ref{sec:light}, using $h^\ast$ in place of $h$. Since $y\notin A$, a chain that reaches $y$ accepts it at the point
$(y,0)$, so each column guess starts the walk from $(y,0)$.

\subsubsection{Analysis}

\paragraph{Capturing every heavy image.}
We first show that $A$ contains every heavy image with probability
at least $1-N^{-2}$. Fix a heavy image $y'$ and let
$\mathrm{Count}_{y'}$ be the number of times it appears among the
$cM/4$ sampled images. Each sample comes from an independent uniform
input and equals $y'$ with probability at least $8/M$, so
$\mathbb{E}[\mathrm{Count}_{y'}]\ge 2c$. Since $c=12\ln N$,
Chernoff's bound gives
$\Pr[\mathrm{Count}_{y'}<c]\le e^{-c/4}=N^{-3}$.
With $M/4$ counters and $cM/4$ samples, Misra-Gries makes at most
$c(M/4)/((M/4)+1)<c$ rounds of simultaneous decrements, so every image
that appears at least $c$ times is retained. A union bound over
at most $N$ heavy images therefore gives probability at most
$N^{-2}$ that $A$ misses any of them. Each counter uses $O(\log\log N)$ bits, so constructing $A$ takes
at most $M$ words of working memory.

\paragraph{Success probability.} For the rest of the analysis we fix
the table $A$ obtained during preprocessing, so all probabilities and
expectations are conditional on $A$, and fix a challenge $y=f(x)$. If $y\in A$, the online algorithm returns its stored
preimage. For $y\notin A$, we give a lower bound on the probability
that $(y,0)$ appears in one of the rainbow tables. We follow the
coverage argument of Section~\ref{sec:light}, using the transition
$h_s^\ast$ and its merging probability $q^\ast$.
As a fixed chain moves from one column to the next, each fresh
function $g_s$ gives it a uniform input to $f$, so it reaches $y$
with probability $I(y)/N$ at each step. Since $y\notin A$, the
chain accepts $y$ whenever it reaches it, and the chain probability of
covering $y$ is again $p_y=1-(1-I(y)/N)^t$. Since $q^\ast$ is the per-column merging probability for two chains, the second-moment argument of Section~\ref{sec:light} carries over with $q^\ast$ in place of $q$, giving the same lower bound on the
probability that a single table covers $y$.
With $m=\lceil1/(tq^\ast)\rceil$, the single-table coverage argument of
Section~\ref{sec:light} carries over with $q^\ast$ in place of
$q$. Conditional on $A$, the $\ell=\lceil 2Nq^\ast+1\rceil$
tables are independent, so the same amplification argument gives
probability at least $3/5$ that at least one table covers $y$.

\paragraph{Time.}
If $y\in A$, we return its stored preimage without evaluating $f$.
Otherwise, each column costs one evaluation of $f$, including when
the step is rejected. Since we compare both the image and the retry
counter at the endpoint, a false alarm still occurs only when an incorrect
candidate chain merges into a stored chain. The
time analysis of Section~\ref{sec:light} therefore carries over with
$q^\ast$ in place of $q$, giving 
\[
\mathbb E[T]\le \ell t^2.
\]

\paragraph{Memory.}
Each chain stores its starting point and its full endpoint $(y_t,a_t)$.
The starting point and the endpoint image together take $2$ words, and
the retry counter takes at most one more, so each chain requires at
most $3$ words. Each column function $g_s$ uses $3$ words
(see Section~\ref{sec:column-functions}), so the $t$ functions use
$3t$ words per table. Finally, $A$ stores at most $M/4$
image-preimage pairs of $2$ words each, that is, $M/2$ words, and we
choose $t$ so that the rainbow tables take the remaining $M/2$ words.
Using $t\le m$ as before, this choice gives
\[
\frac{M}{2}=3\ell m+3\ell t\le 6\ell m,
\]
so $M\le12\ell m$.

\paragraph{Time-memory tradeoff.}
Combining $T\le\ell t^2$ with $M\le 12\ell m$ and using
$mt=O(1/q^\ast)$ gives
\[
TM^2=O(m^2\ell^3t^2)= O\left((mt)^2N^3(q^\ast)^3\right)
     = O(N^3q^\ast),
\]
using $\ell=O(Nq^\ast)$ for the first inequality and $mt=O(1/q^\ast)$ for the second. 

With probability at least $1-N^{-2}$, $A$ contains every heavy image,
so each $z\notin A$ has $I(z)<8N/M$ and
\[
q^\ast=\sum_{z\notin A}\Bigl(\frac{I(z)}{N}\Bigr)^2
<\frac{8}{M}\sum_{z\notin A}\frac{I(z)}{N}\le\frac{8}{M},
\]
so the bound $TM^2=O(N^3q^\ast)$ gives the worst-case tradeoff
$TM^3=O(N^3)$.

\section{Prism Tables}\label{sec:prism}

We first describe and analyze prism tables for the basic construction
of Section~\ref{sec:light}, where the chain points are simply the images
of $f$. Both the construction and its analysis
extend naturally to the full scheme of Section~\ref{sec:heavy}, where
each checkpoint filter stores the pairs $(y,a)$ as the chain points. 

\paragraph{The Algorithm.}
We assume $t\ge7\log N$ and choose $\log N$ (nearly) evenly spaced
checkpoints, with the $s$-th checkpoint at column
\[
c_s \coloneqq \left\lfloor\frac{st}{\log N}\right\rfloor,
\qquad s\in[\log N],
\]
so consecutive checkpoints are at most
$\lceil t/\log N\rceil$
columns apart.%
\footnote{\label{fn:small-t}If $t<7\log N$, we place a filter at every column, so
Lemma~\ref{lem:prism-time} with $d=1$ shows that a wrong guess costs at
most $7$ evaluations in expectation, giving $T=O(\ell t)$. The fewer than
$7\log N$ filters per table take $O(m)$ words, so $M=O(\ell m)$ and
$TM=O(N^2q)$. For a random function this is $TM=O(N)$, which is optimal
up to constants by Yao's bound~\cite{Yao90}.}
During preprocessing, we place a Bloom filter $B_s$ at each
checkpoint $c_s$ of each table, each with false-positive probability at
most $1/2$ and no false negatives. We then build the chains column
by column as before, and whenever we reach a checkpoint $c_s$, we
insert the image $y_{c_s}^j$ of each chain $j\in[m]$
into $B_s$.

When the online algorithm guesses that $y$ appears in column $r$, it
starts from $y$ and develops the candidate chain as before, but now,
whenever it reaches a checkpoint column $c_s\ge r$, it queries $B_s$
with the current image $y_{c_s}$.
A negative answer rules out the guess, while a positive
answer requires the algorithm to continue developing the
chain for at most $\lceil t/\log N\rceil$ additional columns to reach checkpoint
$c_{s+1}$, where the test is repeated. If the candidate chain passes all checkpoints and reaches the endpoints at column $t$, the algorithm
proceeds as in the basic construction: it checks the stored endpoints
and reconstructs any matching chains, moving on to the next guess if
there is no matching endpoint or if none of the matching chains contains
$y$ in column $r$ (Algorithm~\ref{alg:online-prism}).

\subsection{Analysis}
\label{sec:analysis}

\begin{theorem}[Prism tables]\label{thm:prism}
Let $f\colon[N]\to[N]$ and $t\ge 7\log N$ (for smaller $t$, see
footnote~\ref{fn:small-t}).
Adding to the rainbow tables of Section~\ref{sec:heavy} a Bloom filter
at each of the $\log N$ checkpoint columns $c_s$, chosen independently
with false-positive probability at most $1/2$, gives prism tables that
invert each fixed challenge $y=f(x)$ with probability at least $3/5$
and achieve the tradeoff
\[
TM^2=O\left(\frac{N^3q_A^\ast(f)}{\log N}\right).
\]
With probability at least $1-N^{-2}$ over the construction of $A$,
the worst-case tradeoff becomes
\[
TM^3=O\left(\frac{N^3}{\log N}\right).
\]
\end{theorem}

\begin{lemma}\label{lem:prism-time}
Let $\mathrm{Cost}(r)$ be the number of evaluations of
$f$ used to reject an incorrect column guess $r$ (including
false alarms).  If $m=\lceil{1/(tq)}\rceil$ and consecutive checkpoints are at most $d$ columns apart, then
\[
\mathbb{E}[\mathrm{Cost}(r)]\le 7d.
\]
\end{lemma}
\begin{proof}
Fix an incorrect column guess $r$. We separate the work
up to the first merge or rejection, either at a checkpoint
or at an unmatched endpoint, from the additional work
caused by false alarms after a merge.

Before the candidate chain merges with a stored chain, its reached image
at each checkpoint is absent from the corresponding stored
set, so passing the checkpoint requires a false positive of the Bloom filter.
The filters' hash functions are independent of each other and of the
column functions, so passing $k$ checkpoints
before a merge has probability at most $2^{-k}$, and each
checkpoint interval contains at most $d$ steps. Thus the expected number of steps until a merge or termination is at most
\[
d \sum_{k\ge0}2^{-k}
\le 2d.
\]
A merge creates at least one false alarm, and the remaining
walk and chain reconstructions cost at most $t$ additional
evaluations per false alarm. Adding the expected cost of
this remaining work, which is zero if no merge occurs, gives
\[
\mathbb{E}[\mathrm{Cost}(r)]
\le 2d
+t\,\mathbb{E}[\text{number of false alarms}].
\]

We count separately the false alarms created at the first
merge and those created later, since checkpoints may reject
the candidate before its first merge, while afterward it
follows a stored chain and passes every remaining checkpoint.

Conditional on reaching a step without an earlier merge or rejection,
the fresh column function makes the candidate meet each stored chain
with probability $q$, giving $mq$ chains met in expectation, a count
that includes stored chains that merged with each other earlier.
Weighting this by the probability of reaching each step gives
\[
\mathbb{E}[\text{false alarms at the first merge}]
=mq\sum_{s=r+1}^t
\Pr[\text{reaching column }s\text{ with no earlier merge}]
\le 2mqd,
\]
where the sum is the expected number of steps up to rejection,
the first merge, or the endpoint, including the merging step,
which we gave an upper bound above of $2d$. 

After the first merge, each remaining stored chain can meet the
candidate chain in at most $t$ later columns, with probability $q$ per
column. This gives at most $mtq$
additional false alarms in expectation conditional on a first merge.
Weighting this by the probability of a first merge and adding the
chains met at that merge gives
\[
\mathbb{E}[\text{number of false alarms}]
\le 2mqd
   + 2mqd\cdot mtq
\le \frac{5d}{t},
\]
where the last inequality uses $mtq\le 1+1/t$.
Substituting this into the cost bound gives
\[
\mathbb{E}[\mathrm{Cost}(r)]\le 2d+t\cdot\frac{5d}{t}=7d.
\]

\end{proof}

\begin{proof}[Proof of Theorem~\ref{thm:prism}]
We first show that the filters do not reduce the success probability.
Prism tables store the same chains as the rainbow tables they are built
on, and change only the online walks. For a correct column guess, the candidate
walk from $y$ follows the stored chain that contains $y$, so every
filter it queries contains its point and, since Bloom filters have no
false negatives, accepts it. The walk therefore reaches the endpoint,
where the algorithm proceeds as for rainbow tables. Prism tables thus succeed exactly when the underlying rainbow tables do,
which by Theorem~\ref{thm:rainbow}, applied to those tables, happens
with probability at least $3/5$.

\paragraph{Time.} By Lemma~\ref{lem:prism-time} with
$d=\lceil t/\log N\rceil$, and since $t\ge7\log N$, a fixed incorrect guess
costs at most $7d\le8t/\log N$ evaluations in expectation. The online
search tests at most $\ell t$ incorrect guesses, so by linearity of
expectation their total expected cost is at most $8\ell t^2/\log N$.
A correct guess costs at most $t$ evaluations per reconstructed chain,
and the expected number of reconstructions (at the true endpoint) is at most $1+mtq\le3$,
where the first term accounts for the chain that yields a preimage.
Thus we obtain
\[
\mathbb{E}[T]\le \frac{8\ell t^2}{\log N}+3t
\le \frac{9\ell t^2}{\log N}.
\]

As in Section~\ref{sec:light}, counting the evaluations of column
functions, the endpoint lookups and the checkpoint queries as well, the
time remains $O(\ell t^2/\log N)$, since each guess makes only $O(1)$
lookups and queries in expectation.

\paragraph{Memory.}
The underlying rainbow tables use at most $4\ell m$ words, including
the column functions. Each table has at most $\log N$ filters, each
storing at most $m$ elements, so they use $2m\log N$ bits, that is
$2m$ words, plus $2\log N$ words for the hashes. Since
$\log N\le t\le m$, the total memory is
\[
M\le 4\ell m+2\ell m+2\ell\log N\le 7\ell m.
\]

\paragraph{Time-memory tradeoff.}
The memory is $M=O(\ell m)$ words, as for rainbow tables, while the
expected online time $T=O(\ell t^2/\log N)$ is smaller by a factor of
$\log N$, so the tradeoff of Section~\ref{sec:light} improves by the
same factor:
\[
TM^2=O\!\left(\frac{N^3q}{\log N}\right).
\]

\paragraph{Prism tables on the full construction.}
Prism tables apply to the full construction of Section~\ref{sec:heavy}
without change, with each filter now storing the full chain points $(y,a)$ of
its column. Storing the full points instead of images does not change the size of
a filter, which depends only on the number of elements it stores.
Thus as the Memory paragraph above shows, adding $\log N$
filters per table at most doubles the memory of the rainbow tables, so
we give $M/3$ words to the table $A$, $M/3$ to the rainbow tables and $M/3$ to
the filters. Applying Section~\ref{sec:heavy} with $2M/3$ in place of
$M$, the analysis above gives
\[
TM^2=O\!\left(\frac{N^3q^\ast}{\log N}\right),
\]
and since $q^\ast<12/M$ with probability at least $1-N^{-2}$, this
gives the worst-case tradeoff
\[
TM^3=O\!\left(\frac{N^3}{\log N}\right)
\]
of Theorem~\ref{thm:prism}.
\end{proof}

\begin{algorithm}
\caption{Online inversion with prism tables}
\label{alg:online-prism}
\KwIn{$f$, $y\in[N]$, and the tables $\mathcal{T}_1,\ldots,\mathcal{T}_\ell$ with their filters}
\For{$r=t,t-1,\ldots,1$}{
    \For{$i=1,\ldots,\ell$}{
        Let $s$ be the smallest index with $c_s\ge r$\;
        $z\gets (h_{c_s}^{(i)}\circ\cdots\circ h_{r+1}^{(i)})(y)$\;
        \While{$s<\log N$ and $B_s^{(i)}$ accepts $z$}{
            $z\gets (h_{c_{s+1}}^{(i)}\circ\cdots\circ h_{c_s+1}^{(i)})(z)$\;
            $s\gets s+1$\;
        }
        \If(\tcp*[h]{$z$ is in column $c_{\log N}=t$}){$s=\log N$}{
            \ForEach{chain $C_j$ in $\mathcal{T}_i$ with endpoint $z$}{
                $u\gets (h_{r-1}^{(i)}\circ\cdots\circ h_1^{(i)})(y_0^j)$\;
                \lIf{$h_r^{(i)}(u)=y$}{\Return{$g_r^{(i)}(u)$}}
            }
        }
    }
}
\Return{$\bot$}\;
\end{algorithm}
\FloatBarrier

\section{Experiments}\label{sec:experiments}
Our analysis shows that prism tables save a factor of $\log N$ in time,
but it is unclear whether the saving is visible at practical sizes. We
therefore implemented both schemes and compared them on the same
functions, with the same memory and similar success probability.

\paragraph{Functions.}
Although our construction does not require $N$ to be prime, for the
experiments we chose $N$ to be the largest prime below $2^n$, for each
$n\in\{20,22,\ldots,32\}$. This allows the column functions $g_s(y)=\alpha y+\beta$ to be
evaluated using ordinary multiplication and addition modulo $N$. We accordingly defined the functions to be inverted on
the same domain, taking ten functions
$f(x)=\mathrm{AES}_K(x)\bmod N$, one for each of ten keys $K$, where
$\mathrm{AES}_K(x)$ denotes the low $64$ bits of AES-128 under key $K$.
Both schemes use the same functions $f$ and the same family of
column functions.
We used these AES-based functions as a model of random functions and
applied the basic construction of Section~\ref{sec:light}, with one
table for each scheme.

\paragraph{Memory and parameters.}
We count memory in bits and charge each chain $2n$ bits, for its start
and its endpoint, and we do not count the column functions and filter keys, which take far fewer bits than the chains. The rainbow table has $m_{\mathrm{Rb}}$ chains of
length $t_{\mathrm{Rb}}=N^{1/3}$, with $m_{\mathrm{Rb}}=N/t_{\mathrm{Rb}}$,
so its memory is $S=2nm_{\mathrm{Rb}}$ bits, and this is the budget for
both schemes. A prism chain costs the same $2n$
bits, plus one point in each of the $n$ filters, and a Bloom filter
with false-positive probability $1/2$ needs about $1/\ln2\approx1.44$
bits per point, so a prism chain costs $2n+1.44n=3.44n$ bits. 
To stay within the memory of the rainbow table, the prism table must
hold fewer chains, $m_{\mathrm{Pm}}=2/3.44\cdot m_{\mathrm{Rb}}\approx0.58\,m_{\mathrm{Rb}}$,
and we make them longer,
$t_{\mathrm{Pm}}=N/m_{\mathrm{Pm}}\approx1.72\,t_{\mathrm{Rb}}$, so
that both tables have $mt=N$ and hence about the same success
probability. 

The $n$ filters sit at the evenly spaced columns
$\lfloor jt_{\mathrm{Pm}}/(n+1)\rfloor$ for $j\in[n]$, and each uses
AES under its own key as its hash function.

\paragraph{Measurements.}
For each function, both schemes inverted the same $2{,}000$ targets
$y=f(x)$, with $x$ uniform in $[N]$. The rainbow table tries the
cheapest guesses first, $r=t_{\mathrm{Rb}},\ldots,1$, and our prism
implementation does the same with its checkpoints, trying first the
guesses nearest the next checkpoint rather than the order of
Algorithm~\ref{alg:online-prism}. We count every evaluation of $f$ in
the online phase, including rebuilds after false alarms, over all
targets, found or not, and measure CPU time on one core of an Apple
M1 Pro. For each function, we timed each target once per scheme using
\texttt{clock()}, running the rainbow lookups before the prism lookups.

\begin{figure}[htbp]
  \centering
  \includegraphics[width=0.55\linewidth]{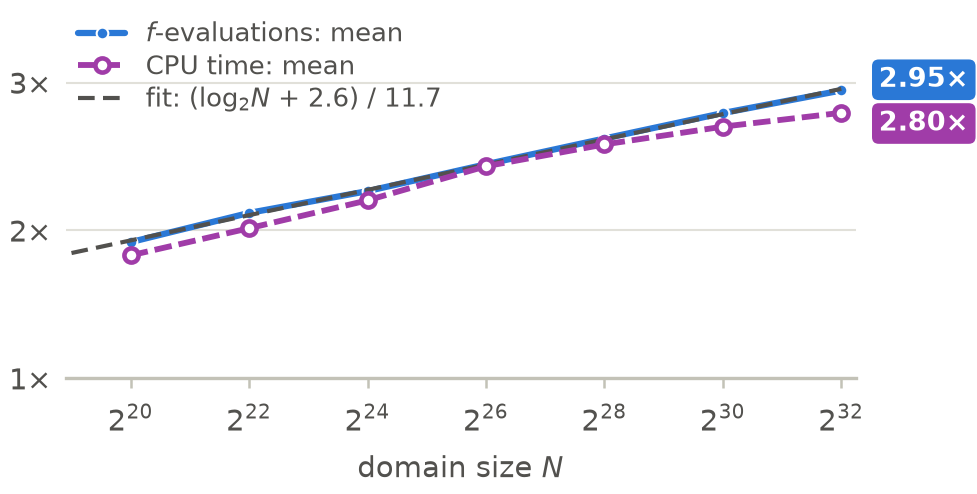}
  \caption{Speedup of prism tables over rainbow tables with the same
  storage and similar success probability, in evaluations of $f$ and in CPU
  time.}
  \label{fig:speedup}
\end{figure}

\paragraph{Results.}
Both schemes found about $74\%$ of the targets at every $n$, and prism
tables used $1.92$--$2.95\times$ fewer evaluations of $f$ and
$1.83$--$2.80\times$ less CPU time (Table~\ref{tab:results}), with a
gain that grows like $(\log_2N+2.6)/11.7$ (Fig.~\ref{fig:speedup}).
Across the ten functions, the standard deviation of the speedup in
evaluations was between $0.03$ and $0.06$ for every $n$.

Intuitively, the gain is about $n/12$ rather than $n$ for two reasons.
First, the prism chains are $1.72$ times longer, and since the online
time grows like the square of the chain length, this costs a factor of
$1.72^2\approx3$. Second, a wrong guess in a rainbow table
walks on average about half a chain, $t/2$ steps, before it is rejected. 
In a prism table, since each checkpoint lets a wrong guess pass with
probability $1/2$, it is rejected on average at the second checkpoint
it reaches, and since an average guess starts halfway between two
checkpoints, it walks about one and a half intervals, or $1.5t/n$
steps. This saves a factor
of $n/3$ rather than $n$, so together the two effects give about
$n/9$. The measured gain is a little smaller because in both tables a
successful search stops on average halfway, and this saves the
rainbow table more, since its cheapest guesses come first, while the
guesses of a prism table all cost about the same.

\begin{table}[htbp]
\centering
\small
\setlength{\tabcolsep}{4pt}
\begin{tabular}{r rr rr cc rr cc}
\toprule
 & \multicolumn{2}{c}{Rainbow} & \multicolumn{2}{c}{Prism}
 & \multicolumn{2}{c}{Success (\%)}
 & \multicolumn{2}{c}{Mean evaluations of $f$}
 & \multicolumn{2}{c}{Speedup} \\
\cmidrule(lr){2-3}\cmidrule(lr){4-5}\cmidrule(lr){6-7}%
\cmidrule(lr){8-9}\cmidrule(lr){10-11}
$n$ & $m$ & $t$ & $m$ & $t$ & Rainbow & Prism & Rainbow & Prism
 & evals & CPU \\
\midrule
20 & 10{,}280 & 102 & 5{,}971 & 176 & 74.8 & 74.4 & 3{,}182 & 1{,}656 & 1.92 & 1.83 \\
22 & 26{,}052 & 161 & 15{,}134 & 277 & 74.3 & 74.5 & 7{,}918 & 3{,}737 & 2.12 & 2.02 \\
24 & 65{,}536 & 256 & 38{,}072 & 441 & 74.7 & 74.1 & 19{,}833 & 8{,}751 & 2.27 & 2.20 \\
26 & 165{,}293 & 406 & 96{,}025 & 699 & 74.5 & 74.3 & 50{,}135 & 20{,}482 & 2.45 & 2.43 \\
28 & 416{,}179 & 645 & 241{,}774 & 1{,}110 & 74.0 & 74.3 & 127{,}510 & 48{,}630 & 2.62 & 2.58 \\
30 & 1{,}048{,}576 & 1{,}024 & 609{,}159 & 1{,}763 & 74.5 & 74.7 & 317{,}961 & 113{,}793 & 2.79 & 2.70 \\
32 & 2{,}643{,}057 & 1{,}625 & 1{,}535{,}457 & 2{,}797 & 74.3 & 74.1 & 803{,}851 & 272{,}765 & 2.95 & 2.80 \\
\bottomrule
\end{tabular}
\caption{Rainbow and prism tables with the same storage, for $N$ the
largest prime below $2^n$, over ten keys and 2{,}000 targets per key.
Prism tables use $n$ Bloom filters with false-positive probability $1/2$.
Speedups are ratios of means, which include failed searches.}
\label{tab:results}
\end{table}

\FloatBarrier
\section*{AI Methodology} The constructions of prism tables and rigorous rainbow tables, together with their analysis, were developed by the authors over the past few years. AI tools, specifically Claude Opus 5.5 (Extra) and GPT-6 Astra (Extra High), were used to run the experiments and for comments on the writing.

\section*{Further Research}
The obvious question is whether the time-memory tradeoffs can be improved. In particular, whether recent developments regarding the better than exponential algorithm for $k$-sat (see for instance~\cite{cmu26}) can be used here. 

Another issue is whether it is possible to translate the framework described in this work to obtain better circuits for inverting a function, as the work of Mazor and Pass~\cite{MazorPass2024} and Hirahara, Ilango, and Williams~\cite{HIW24} require. 

\bibliographystyle{plainurl}
\bibliography{references}

\end{document}